# Chalcogen Height Dependence of Magnetism and Fermiology in $FeTe_xSe_{1-x}$

Jagdish Kumar[1,2], S. Auluck[1], P.K. Ahluwalia[2] and V.P.S. Awana[1]

[1]*Quantum Phenomena and Application Division, National Physical Laboratory, New Delhi-110012, India*

[2]*Department of Physics, Himachal Pradesh University, Shimla-171005, India*

$FeTe_xSe_{1-x}$ (x=0, 0.25, 0.50, 0.75 and 1) system has been studied using density functional theory. Our results show that for FeSe, LDA seems better approximation in terms of magnitude of magnetic energy whereas GGA overestimates it largely. On the other hand for FeTe, GGA is better approximation that gives experimentally observed magnetic state. It has been shown that the height of chalcogen atoms above Fe layers has significant effect on band structure, electronic density of states (DOS) at Fermi level $N(E_F)$ and Fermi surfaces. For FeSe the value of $N(E_F)$ is small so as to satisfy Stoner criteria for ferromagnetism, ($I \times N(E_F) \geq 1$) whereas for FeTe, since the value of $N(E_F)$ is large, the same is close to be satisfied. Force minimization done for $FeTe_xSe_{1-x}$ using supercell approach shows that in disordered system Se and Te do not share same site and have two distinct $z$ coordinates. This has small effect on magnetic energy but no significant difference in band structure and DOS near $E_F$ when calculated using either relaxed or average value of $z$ for chalcogen atoms. Thus substitution of Se at Te site decreases average value of chalcogen height above Fe layers which in turn affect the magnetism and Fermiology in the system. By using coherent-potential approximation for disordered system we found that height of chalcogen atoms above Fe layer rather than chalcogen species or disorder in the anion planes, affect magnetism and shape of Fermi surfaces (FS), thus significantly altering nesting conditions, which govern antiferromagnetic spin fluctuations in the system.

PACS: 71.15.Mb, 74.70.Xa, 75.10.Lp, 74.20.Mn, 74.25.Ha, 71.18.+y

## 1 Introduction

The iron chalcogenides belong to an interesting class of ferrous superconductors. These materials consist of stacked layers of Fe-X (X=S, Se and Te) along *c*-axis which play a key role in superconductivity [1]. In contrast to iron pnictides, they do not have charge reservoir layers in addition to superconducting Fe-X layers and thus have an advantage of structural simplicity. For iron chalcogenides, parent compound FeTe is non superconducting and have long range "double-stripe" (DS) antiferromagnetic (AFM) ordering [2], [3] and [4]. This double stripe AFM ordering is different from the iron pnictides which exhibit "single stripe" ordering [5], [6] and [7]. Unlike pnictides, FeTe does not require doping by external charge carriers (electrons/holes) for appearance of superconductivity. However, substitution by isoelectronic Se/S at Te site is known to suppress long range antiferromagnetic ordering in FeTe and induce superconductivity with maximum transition temperature ($T_c$) of ~14K for $FeSe_{0.5}Te_{0.5}$ [8]. FeSe does not have any long range magnetic ordering and has superconducting transition temperature around 8K [1] which increases dramatically to 37K under pressure of ~8GPa [9], [10], [11], [12] and [13]. These facts underline the crucial role of structure on mechanisms of superconductivity in these compounds. Moreover, both magnetism and superconductivity in iron chalcogenides are very sensitive to stoichiometry and content of interstitial iron [14], [15], [16], [17] and [18]. While perfectly stoichiometric FeSe is nonmagnetic and superconducting [14], any deficiency or excess of Fe results in ferri/ferromagnetism [15], [16].

Theoretical calculations by Kuroki et al. [19] have shown that height of anion atoms above Fe layers could be acting as a switch between high-$T_c$ nodeless and low-$T_c$ nodal pairings for Fe based superconductors. Based on this idea Mizuguchi et al. [20] using experimental data have found interesting relationship between superconductivity and height of anion atoms above Fe layers for different classes of Fe based superconductors. They found that for optimally doped compounds, the superconducting transition temperature $T_c$ shows a peak for height value of 1.38Å, above and below which the value of $T_c$ decreases sharply. Similar dependence was found by Okabe et al [21] for FeSe under high pressure studies. Based on DFT calculations Kuchinskii et al. [22] have found that variation of $T_c$ as function of anion height is due to the corresponding change in the density of states at the Fermi level and can be explained semi-quantitatively using BCS theory. They also suggested possibility of non-phonon pairing in region of maximum $T_c$. Density functional calculations by C. Y. Moon et al. [23] using pseudopotential based method also supports that height of chalcogen atoms above Fe layers decides magnetism in chalcogenides. Also, Moon et al. have used only generalized gradient approximation (GGA) in their study. However, we have used both local density approximation (LDA) and GGA and marked the difference between the two approximations for studied system. Thus, height of anion atoms above Fe layers is of crucial importance for all ferrous superconductors which is being further probed in this paper for iron chalcogenides by using more accurate all electron full potential linearized augmented plane wave method (FP-LAPW).

In DFT calculations magnetic ground state for FeSe is found to be antiferromagnetic with stabilization energy of ~5meV/Fe [24]. As a general experience, magnetism in metals near quantum critical points can be entirely suppressed by spin fluctuations if the LDA magnetic energy is of the order of 10-15 meV/atom, but it is rather hard for spin fluctuations to entirely destroy magnetism that is stabilized by any substantially larger energy [25]. One of the dominant mechanisms behind spin fluctuations for these materials is nesting of electron and hole pockets at Fermi surface (FS), which are further thought to be a mechanism behind superconducting pairing of electrons [26].

Gradient corrections through generalized gradient approximation (GGA) to the LDA generally results in overestimated magnetic energies. For itinerant or near itinerant magnets, LDA results are closer to experiments in terms of magnetism. On the other hand, GGA is usually better at predicting the crystal structure of transition metal compounds [24], [25]. Thus it is reasonable to relax the atomic positions using GGA and then calculate magnetic energies using LDA [25]. Though, for FeSe use of both LDA and GGA for calculation of magnetic properties can be found in literature but for FeTe all the reports have used GGA without mention of its LDA counterpart [2], [16], [23], [24] and [27]. Therefore, in this paper we have reported detailed calculations on magnetism of these systems by using one of the most accurate first principle technique (FPLAPW+lo) based methods and found that both LDA and GGA give "single stripe" (SS) as the ground state for FeSe which is consistent with earlier reports [2], [23] and [24]. However, for FeTe, GGA gives DS as the ground state in agreement with earlier calculations [2], [23] and experiments [3], [4], but LDA gives SS as the ground state which is contrary to experimental results.

We have also studied the effect of Se disorder at Te site in FeTe$_{1-x}$Se$_x$. First the structural properties of disordered system are studied using supercell approach within FPLAPW method, where it is found that Se and Te atoms relax at two different heights above Fe layers and these values remain almost same throughout the composition. These two distinct *z* values of Se and Te atoms slightly enhance the spin fluctuations in the system in comparison to average *z* value for both Se and Te. However, electronic bands and DOS near FS remain practically same for relaxed and average *z* taken for Se and Te.

Further, the electronic properties of disordered system FeTe$_x$Se$_{1-x}$ are looked at by using SPRKKR code based on coherent potential approximation (CPA), which offers a reliable approximation to describe the effects of disorder in these systems [28], [29]. We have found that there is a decrease in average value of chalcogen height with increase of Se content in FeTe$_x$Se$_{1-x}$. It is observed that height of anion atoms (rather than disorder) above Fe layers act as crucial parameter tailoring magnetic and electronic properties in this system.

## 2 Computational Details

All the calculations have been performed using full potential linear augmented plane wave plus local orbital (FPLAPW+lo) method as implemented in Wien2k code [30]. The muffin tin radii for Fe and Se is chosen 2.00 a.u. and for Te 2.10 a.u. The calculations of total energies in different magnetic states were performed for more than 100 k-points in irreducible Brillouin Zone (IBZ). As stated earlier the magnetic ground state for FeTe is sensitive to the exchange-correlation used so the contribution to the electron-electron interaction has been modelled using LDA [31] as well as the GGA [32]. For the wave function expansion inside the atomic spheres, a maximum value of $l_{max}$=12 is used, together with a plane-wave cut off of $R_{MT}K_{max}$=7.0 with $G_{max}$=24. We have ensured convergence of total energy with different computational parameters such as number of *k*-points, number of plane waves chosen and maximum value of angular momentum.

At room temperature both FeSe and FeTe crystallize in tetragonal structure with spacegroup *P4/nmm* (Spacegroup no. 129). Fe is located at 2a site (0.75, 0.25, 0.00) and Se/Te at 2c site (0.25, 0.25, *z*). The experimental values of lattice parameters for FeSe [1], FeSe$_{0.25}$Te$_{0.75}$ [33], FeSe$_{0.5}$Te$_{0.5}$ [34], FeSe$_{0.75}$Te$_{0.25}$ [33] and FeTe [35] were taken. The term chalcogen height ($h_{se}$ and $h_{Te}$) corresponds to height of chalcogen atoms above Fe layer and refers to $h=z\times c$, where *c* is length of ***c*** vector chosen in the direction of *c*-axis. The Fe-X bond distance is given by $\sqrt{a^2 + 4h^2}/2$ and X-Fe-X tetrahedral angle by $2\cot^{-1}\left(\frac{2h}{a}\right)$, where *a*, is lattice parameter of tetragonal cell. The value of '*a*' changes by only ~1.5 percent and *h* changes by almost 30 percent while going from FeSe to FeTe. Therefore, chalcogen height above Fe layers can be considered as a key parameter in deciding Fe-X bonding and tetrahedral angles. The electronic properties and magnetism in these iron based superconductors is found very sensitive to the Fe-X bond distances [25], [36] and tetrahedral angles [9], [37]. Many authors have discussed this dependence in terms of height of anion atoms (*h*) above Fe layers [19], [20], [21], [22] [23] and [37]. Thus we shall discuss our results in terms of chalcogen height so as to have direct comparison with earlier results. The variation of chalcogen height is achieved by varying *z*, while keeping lattice parameters fixed to their experimental values.

Different magnetic states considered are nonmagnetic (NM), ferromagnetic (FM) and three different antiferromagnetic (AFM) states identified as single stripes (SS), double stripes (DS) and checkerboard (CH). The NM and FM states retain the crystal symmetry *P4/nmm* (Spacegroup No. 129) with 2 formula units per cell (f.u./cell). Checkerboard (CH) type antiferromagnetic state has *P-4m2* symmetry (Spacegroup No. 115) with 2 f.u./cell. Single stripes (SS) consists of alternating stripes along 100 direction having *Pccm* symmetry (Spacegroup No. 49) with 4 f.u./cell and double stripes (DS) with *P2$_1$/m* symmetry (Spacegroup No. 11) have 4 f.u./cell. These different magnetic states are illustrated in Figure 1(a)-(c).

The optimized value for height of chalcogen atoms are obtained by using force minimization technique to a force tolerance of 0.5mRy/a.u. For FeSe we did force minimization in NM state whereas for FeTe we used DS state in accordance with experiments. The optimization done by including magnetism, results in better agreement between calculated and experimental atomic positions, Table 1. The structural properties of disordered system are studied using supercell approach using WIEN2k code.

Since, CPA has been found to reliably describe the effects of disorder in these systems [28], [29] so we have also studied the effect of Se disorder on electronic properties of FeTe$_x$Se$_{1-x}$ system by employing Korringa-Kohn-Rostoker coherent potential approximation (KKR-CPA) method in atomic sphere approximation (ASA) as implemented in SPRKKR code [38]. Our CPA calculations were performed using the tetragonal structure with lattice parameters from experiment. Additionally, in these calculations we used atomic positions obtained after optimisation from Wien2k. The idea of the CPA is to replace the random array of real on-site potentials by an ordered array of effective potentials. The scattering properties of the effective potential are then determined self-consistently in terms of the local mean-field theory with the condition that the total Green's function of the effective system does not change upon replacement of the single effective potential with the real one. To reduce the errors due to the ASA, we have introduced four empty spheres in the unit cell containing two Fe and two Se atoms. Here, we have used local density approximation by Barth and Hedin [39] for exchange and correlation effects. The Brillouin zone (BZ) integration during self-consistency was carried out using a grid of 18×18×12 points containing 385 k-points in IBZ. For both DOS and Bloch spectral function (BSF) calculations, we have added a small imaginary component of 1 *mRyd* for broadening the energy.

## 3 Role of Chalcogen height on magnetic and electronic properties of FeSe and FeTe

As described above the height of chalcogen atoms above Fe layers is one of the crucial parameters for various properties of these systems. Also while calculating the magnetic ground state for optimized or experimental value of $h_{Te}$ for FeTe, we found that, LDA and GGA results in two different magnetic ground states i.e. SS and DS respectively. To investigate the detailed reason behind this kind of behavior we have systematically investigated various magnetic states for different values of $h_{Te}$ using LDA and GGA exchange correlations. Detailed results on magnetism are described in section 3a. In section 3b we have examined chalcogen height dependence on electronic properties of FeSe.

### 3a. Magnetism

Magnetic energy and magnetic moment on Fe as a function of height ($h$) of chalcogen above Fe layers calculated using LDA [31] and GGA [32] for FeSe is shown in Figures 2 (a) and (b) respectively. Figures 3(a) and (b) depict corresponding plots for FeTe. The optimized and experimental values of $h_{Se}$ for FeSe are marked by dashed vertical lines $h_{opt}$ and $h_{expt}$ respectively in Figure 2. For comparison, we have used LDA and GGA exchange correlation. Some of the common features of these systems are as follows.

For SS state Fe retains its magnetic moment up to much lower height of chalcogen atoms compared to other magnetic states. Below this value of chalcogen height there is no magnetic moment on Fe atoms and thus all the states are energetically degenerate to nonmagnetic state. This is due to stronger hybridization of Fe-$d$ states with $p$ states of chalcogen atoms. As the height of chalcogen atoms increase above Fe layer there is a cross over from NM to SS state, SS to DS state and finally from DS to FM state. This feature is qualitatively common for both FeSe and FeTe irrespective of the choice of exchange correlation and is consistent with earlier reports [23]. As can be seen from Figures 2 and 3, SS state is ground state over a large range of chalcogen height. In case of FeSe, SS is the ground state for experimental and optimized value of $h_{Se}$. However, for experimental value of $h_{Se}$, stabilization energy ($E-E_{NM}$) of SS is above ~100*meV/*Fe for both LDA and GGA which is quite large. For optimized value the stabilization energy of SS state is 24*meV/*Fe in GGA and 5*meV/*Fe for LDA.

The dynamical effects like spin fluctuations can suppress the magnetic ordering of the states which have stabilization energy of the order of 10-15meV/Fe [25]. Since FeSe does not exhibit any long range magnetic ordering and GGA overestimates magnetic energy, thus one should use LDA to account for no long range magnetic ordering in FeSe. As can be seen from Figure 2 (a) and (b) experimental value of $h_{Te}$ (3.3515Å) comes in region (light grey) where double stripes (DS) state is ground state for "FeSe" in both LDA and GGA.

Looking at the plots in Figure 3 (a) we find that in case of LDA both experimental and optimized value of $h_{Te}$ lie in SS region which is in contrast to experiments [3]. However, in case of GGA (Figure 3(b)) both optimized (done in DS state not in NM) and experimental value of $h_{Te}$ (3.3515Å) lie in DS range which is in accordance to experimental results [3]. Thus, we conclude that one should use GGA for FeTe in order to explain the experimentally observed DS state. Optimized value of $h_{Te}$, lies close to crossover from SS to DS, below which SS is the ground state. Since optimization done in NM state underestimates the value of $h_{Te}$ (see Table 1), thus one finds SS as ground state corresponding to this value of $h_{Te}$. Therefore, despite having same structure with Se and Te being isoelectronic, LDA seems suitable for FeSe but not for FeTe. To describe FeTe one has to use GGA for both structural and magnetic calculations to find agreement with experiments. This is in contrast to iron pnictides and their parent compounds, where LDA gives better results to describe the magnetic energies and ground state [25]. The DS ordering observed in FeTe is rather novel as this state is stable only over a small range of height values which indicates that this state is a result of delicate balance between antiferromagnetic and ferromagnetic tendency of system. Moreover, the value of magnetic moment on Fe is almost saturated and same for all the magnetic states at higher values of $h$. This represents more localized nature of the magnetism in system for higher values of chalcogen height. Thus system moves from an itinerant magnetic state to localized magnetic state as the height of chalcogen atoms is increased above Fe layers. This is in agreement with earlier calculations by Pulikkotil et al. [36]. In fact magnetism in FeTe has been explained using Heisenberg model based on local moment picture [2].

**3b. Electronic density of states (DOS), band structure and Fermi Surface (FS)**

As for magnetism, electronic properties are also quite sensitive to the height of chalcogen atoms above Fe layer. Since optimization done in NM state underestimates the value of $h$ largely whereas incorporating magnetism for optimization, results are in good agreement to calculated and experimental value of *h* (see Table 1). Thus it is reasonable to ask, how important the *h* value is, for calculations of electronic properties. We have thus investigated the electronic properties as function of chalcogen height above Fe layers.

Figure 4(a) shows the value of electronic density of states at Fermi level $N(E_F)$ as a function of chalcogen height for FeSe and FeTe. It can be found that the value of $N(E_F)$ increases as the height of chalcogen atoms increases above Fe-layer. The trend is similar for both FeSe and FeTe except for height below 2.5Å. With increasing the height of chalcogen atoms above Fe layers the bond distance between Fe and chalcogen atoms increases. The main contribution to the $N(E_F)$ comes from Fe-d states. For smaller value of bond distance the Fe-d bands are more delocalized and are distributed mainly from around -3.0eV to +3eV. With increase in chalcogen height and thus bond distance between Fe and chalcogen atoms Fe-d bands becomes more localized and shifts towards Fermi level. Which results in increase in value of $N(E_F)$ as the height of chalcogen atoms above Fe layers increase. The value of $N(E_F)$ is directly proportional to Sommerfeld constant γ determined from electronic specific heat [40].

$$\gamma = \frac{1}{3}N(E_F)\pi^2 K_B^2$$

The value of γ obtained from calculated value of $N(E_F)$ are given in table 2 ($\gamma_{calc.}$). Generally γ measured in experiments is enhanced by a factor of (1+λ) in comparison to calculated value. Where enhancement factor 1+λ corresponds to the phenomena which are not considered in band stricture calculations such as electron phonon coupling, spin fluctuations etc. The value of λ is also given in table 2. The large value of enhancement factor can be interpreted in terms of large mass enhancement $m^*/m_{band}$ for these materials which can be as large as 20 [49]. It can be seen that the value of λ increases as going from FeSe to FeTe$_{0.75}$Se$_{0.25}$ and then decrease for FeTe. Moreover, $\gamma_{calc.}$ increases linearly with increasing *x* in FeSe$_{1-x}$Te$_x$, for atomic positions obtained via atomic relaxation in NM state. Whereas for FeTe, taking $h_{Te}$ obtained from relaxing atomic positions in DS state, $\gamma_{calc.}$ deviates from linearity. The non-linear behaviour of γ obtained in experiments indicates increasing role of magnetism while going from FeSe to FeTe. The significantly larger enhancement factor for FeTe$_{0.75}$Se$_{0.25}$ can be attributed to enhanced spin fluctuations in this system. The decrease in same for FeTe can be attributed to long range AFM ordering and/or to structural phase transition in FeTe [50].

Figure 4(b) shows corresponding value of $I \times N(E_F)$, *I* being Stoner parameter [42] and $N(E_F)$ is electronic density of states per Fe atom at Fermi level for both the spins. The higher value of $N(E_F)$ can lead to the magnetism in the system [19], [27]. As per Stoner criteria a system may exhibit ferromagnetic instability if the value of the product $I \times N(E_F)$ is greater than or equal to 1. It is found that the value of $h_{Se}$ for FeSe is such that $N(E_F)$ is ~1.33states/Fe atom which is too small for satisfying Stoner criteria ($I \times N(E_F)$ ~0.70) for ferromagnetism, Figure 4(b). For FeTe, $h_{Te}$ is quite large and $N(E_F)$ is close to satisfy Stoner condition indicating that system is close to ferromagnetic instability. This is in agreement with the fact that both FeSe and FeTe shows a crossover from DS state to FM state at higher values of chalcogen heights.

Figures 5(a)-(d) show electronic band structure and corresponding FS for FeSe at four different values of Se heights above Fe-layers. Interestingly band structure and shape of FS is quite sensitive to the value of $h_{Se}$. The FS are more two dimensional for lower values of $h_{Se}$ reflecting weak interactions along *c*-axis. As the $h_{Se}$ increases the interaction of FeSe layers along *c*-axis increases resulting in a three dimensional character of the bands. Similar sensitivity for shape of FS and band character is found for FeTe.

In general, the electronic structures of the Fe-based superconductors show disconnected FS consisting of hole sections around the zone centre (along Γ-Z) and two electron sections at the zone corner (along M-A direction). Shape of our calculated FS agrees well with earlier reports [24], [26], [28], and [29].

## 4 Effect of Se disorder in FeTe$_x$Se$_{1-x}$

**4a. Structural and magnetic properties**

We have further studied the effect of Se disorder on structural properties of FeTe$_x$Se$_{1-x}$ (*x=0.0, 0.25, 0.50 0.75* and *1.0*). Atomic relaxation (Wien2k) by force minimization for disordered systems result in Se and Te atoms at two different *z* positions on *2c* site which is in agreement with earlier calculations and experiments [43], [44] and [45]. Table 2 gives lattice parameters and the comparison of relaxed atomic positions obtained. Interestingly, Fe-Te bonds in disordered system become shorter than in the binary FeTe, while the Fe-Se bonds in disordered system stay almost the same as in the binary FeSe. This result is consistent with earlier independent experimental reports based on high resolution XRD on single crystals [43], neutron pair distribution functions (PDF) [44] and extended x-ray absorption fine structure [45]. Moreover, it is interesting to note that Fe-Te and Fe-Se bond distances in disordered system remain almost same in magnitude irrespective of the extent of disorder. The height of chalcogen atoms above Fe layers plays a crucial role in magnetism and electronic properties of this system as already discussed above. Thus, to study the effect of different chalcogen positions, we have investigated magnetic and electronic properties in FeSe$_{0.5}$Te$_{0.5}$ system (using Wien2k) for optimized values of chalcogen heights and also for their average value. In nonmagnetic state, the structure with fully relaxed atoms was found to be stable by 121.72*mRy/fu* in comparison to structure with average value of *z* for both Se and Te.

For magnetic study we have considered SS, DS, FM and NM states as described in section 3. Since, in section 3a we found that for FeSe$_{0.5}$Te$_{0.5}$ system average value of chalcogen height lie in regime of itinerant magnetism where GGA do overestimate magnetic energies. Thus magnetic calculations were performed using LDA. It is found that stabilization energy of SS state is 27.27*meV/fu* i.e. 13.64*meV/*Fe and 25.90*meV/fu* i.e. 12.95*meV/*Fe for relaxed and average value of $z_{Se/Te}$ respectively. DS state is found unstable relative to NM state for both relaxed and average value of $z_{Se/Te}$ by 2.05*meV/fu* i.e. 1.02*meV/*Fe and *1.85meV/fu* i.e. 0.92*meV/*Fe respectively. Thus SS is calculated magnetic ground state for FeSe$_{0.5}$Te$_{0.5}$ system as was found for FeSe (section 3a). It is interesting to note that the stabilization energy for SS in FeSe is 5*meV/*Fe whereas for FeSe$_{0.5}$Te$_{0.5}$ it is ~13*meV/*Fe. Since experimentally no long range magnetic ordering is observed in FeSe or FeSe$_{0.5}$Te$_{0.5}$ thus in comparison to FeSe, FeSe$_{0.5}$Te$_{0.5}$ seems to have enhanced spin fluctuations that might be responsible for suppression of magnetic ordering. With relaxed atomic positions, SS state is 0.7*meV/*Fe more stable than for average value of $z_{Se/Te}$. Thus with Se and Te having different heights, system seems to have enhanced spin fluctuations than for average value of Se/Te height. These facts support the higher value of $T_c$ observed for FeSe$_{0.5}$Te$_{0.5}$ in comparison to FeSe with spin fluctuations driven superconductivity.

**4b. Electronic properties**

The electronic properties of FeTe$_x$Se$_{1-x}$ (*x=0.0, 0.25, 0.50 0.75* and *1.0*) system are calculated using relaxed atomic positions. As discussed above, force minimization result in Se and Te atoms to relax at two different *z* positions. Though, there is a redistribution of states above and below Fermi level, we did not find any significant difference in electronic density of states or band structure near Fermi level, even if we take average value of *z* co-ordinate for both Se and Te. Since, superconducting properties are characterized by electrons near the FS, thus we have calculated the electronic properties using average value of *z* co-ordinate for chalcogen atoms. It is well known that Coherent potential approximation (CPA) based methods are more reliable to study disorder in these systems [28], [29], therefore we have used SPRKKR code [38] to calculate electronic properties and FS of disordered system being studied in this paper. We found a good agreement in electronic density of states and band structure as calculated for FeSe and FeTe with SPRKKR and more accurate full potential based Wien2k code and earlier reports [24], [28] and [29]. Thus, all the electronic properties of disordered systems reported hereafter follow from SPRKKR code. The value of electronic density of states at Fermi level $N(E_F)$ is given in Table 2. It is found that there is monotonic increase in value of $N(E_F)$ as the value of *x* increases. The increase is almost linear except for large value in case of FeTe. The large increase for FeTe is because we have used *DS* magnetic state for relaxed atomic positions in case of FeTe which places Te atoms quite higher above Fe layers in comparison to relaxation done in NM state. As discussed earlier in section 3(b) value of $N(E_F)$ is quite sensitive to the value of this height. In case superconductivity in these materials was of purely electron-phonon type [46], FeSe should have minimum $T_c$ and should increase towards FeTe. However, this is contrary to experiments which show presence of non-phonon type pairing mechanism as has already been speculated in these materials [24].

We have also calculated Bloch spectral function (BSF) which represents *k*-resolved density of states [47] and is equivalent to the band structure in the disordered system. When evaluated in a particular plane in BZ for Fermi energy it represents the FS in that particular plane. Figures 6 (a)-(d) present the BSF calculated for FeTe$_x$Se$_{1-x}$ system for different values of *x*. The calculated BSF for pure FeSe and FeTe are in good agreement with band structure calculated using full potential Wien2k code. It can be seen that mainly the bands that consist of hybridized Fe and Se/Te which lie around ±2eV from Fermi level, get affected due to Se substitution at Te site and bands near Fermi level remains almost intact. With substitution of Se at Te site the average height of chalcogen atoms above Fe layer decreases, Table 2. As discussed in section 3b, the shape of FS is quite sensitive to the height of chalcogen atoms above Fe layers. Therefore, we have calculated the FS of FeTe$_x$Se$_{1-x}$ (x=0.0, 0.5 and 1.0) in Γ-X-M, Z-R-A and Γ-M-A planes in BZ and looked at the nesting effects. To see the effect of disorder and chalcogen species we have also calculated the FS for FeSe and FeTe by taking experimental lattice parameters and optimized atomic positions of FeSe, FeSe$_{0.5}$Te$_{0.5}$ and FeTe. In case of FeSe$_{0.5}$Te$_{0.5}$ we have used average value of optimized atomic positions. The results are depicted in Figure 7(a)-(c) respectively.

We can get some qualitative idea about nesting in the system without calculating actual susceptibility through careful examination of shape of electron and hole pockets. As an example, if the FS around the Γ point matches exactly with the FS around the M point when displaced by a reciprocal space vector then the nesting is optimal [28]. In the present context, the optimal nesting corresponds to FS at (Γ-M) and (Z-A) points having matching radii and sharp FS (reflected by red lines in the Figure 7). Any deviation, either from the matching radii or sharpness of the FS (reflected by diffused and/or broadened lines in the Figures), generally reduces the effect of nesting [28]. Since the mechanism of superconductivity in these systems is speculated in terms of antiferromagnetic spin fluctuations [26] that are driven by FS nesting thus shape of FS is expected to strongly affect the superconducting properties of this system.

It can be seen that size of electron/hole surfaces is almost independent of chemical species or disorder. Figure 7(a) gives FS of FeSe and FeTe calculated by using experimental lattice parameters and optimized atomic positions for FeSe. From Figure 7(a) we can see that eccentricity of the electron sections along M-A direction is larger for FeTe than for FeSe. If their shape is perfectly circular (zero eccentricity) just as the hole pockets are, they would result in optimal nesting conditions. Figure 7(b) shows FS calculated for FeSe, FeTe$_{0.5}$Se$_{0.5}$, and FeTe using experimental lattice parameters and average value of relaxed atomic positions of Se and Te in FeTe$_{0.5}$Se$_{0.5}$. We can see a nice matching in size and shape of electron and hole pockets in both Γ-X-M and Z-R-A planes almost independent of chalcogen species or disorder in chalcogen planes. This represents optimal nesting and thus enhanced spin fluctuations in FeTe$_{0.5}$Se$_{0.5}$. This is consistent with our previous conclusion that for FeTe$_{0.5}$Se$_{0.5}$, *SS* AFM state has larger magnetic energy stability than for FeSe and thus an indication of enhanced spin fluctuations in the system since there is no long range magnetic ordering for FeTe$_{0.5}$Se$_{0.5}$. Looking at Figure 7(c) representing FS calculated for experimental value of lattice parameters and optimized atomic positions of FeTe, we found that shape of electron sections around M-A direction is completely different than hole sections along Γ-Z direction. This results in total loss in nesting effects and thus possibility of spin fluctuations.

We have also calculated the electronic band structure for FeSe for different values of *c*-axis keeping height of Se atoms fixed to optimized value (using Wien2k code), where we did not find any significant change in band structure near Fermi level and FS. Thus the change in FS can primarily be attributed to average height of chalcogen atoms above Fe layers. Thus it can be concluded that disorder have minimal effect on shape of the FS and major change in shape of FS is attributed to crystal structure or mainly to Fe-X bond distances and tetrahedral angles being reflected in height of chalcogen atoms above Fe layers. Moreover, as stated in section 3a, we find that there is increase on moment on Fe and thus magnetic energy of different magnetic states with increasing height of chalcogen atoms above Fe layers. Also there is enhancement in conditions of FS nesting going from FeSe to FeTe$_{0.5}$Se$_{0.5}$, and gets lost in going to FeTe. Thus chalcogen height above Fe layers seems to act as an important parameter that controls magnetism, Fermiology and hence the nesting effects in these systems.

## 5 Conclusions

In conclusions we have investigated the role of height of chalcogen atoms above Fe layers on magnetic and electronic properties. It is found that stability of a particular magnetic state depends strongly upon height of chalcogen atoms above Fe layers. The system moves from an itinerant nature of magnetism to localized nature of magnetism as the height of chalcogen atoms increases above Fe layers. It is found that LDA is better choice for magnetic description of FeSe but not for FeTe as GGA is required for getting magnetic ground state consistent with experiments. The novel DS type ordering observed in FeTe, seems to be an interesting outcome of delicate balance between antiferromagnetic (SS) and ferromagnetic (FM) tendency of system being controlled by Fe-X distance. Electronic band structure, density of states and Fermi surface is found strongly dependent on value of height of anion atoms above Fe-layers. The exchange correlation used does not have much effect on electronic band structure of density of states.

The optimization done in NM state underestimates the value of *h* largely whereas incorporating magnetism for optimization, results are in good agreement to calculated and experimental value of *h*. In disordered system, optimization results in Se and Te occupying different sites that correspond to their different heights above Fe layers. This has small impact on magnetic energies, though bands near Fermi level and thus FS remains practically same.

Substitution of Se at Te site in FeTe, results in decrease of average height of chalcogen atoms above Fe layer. The disorder in chalcogen planes and length of *c*-axis has negligible effect on bands near Fermi level and thus shape of FS as far as height of chalcogen atoms above Fe layers is kept constant. Moreover, varying height of chalcogen atoms above Fe layer affect band structure and FS significantly. Therefore, height of chalcogen atoms above Fe layers play key role on magnetism, shape of FS and thus nesting effects. In case superconductivity is mediated by spin fluctuations in these systems which are driven by FS nesting, height of chalcogen atoms above Fe layers act as a switching parameter between magnetism and superconductivity.

## 6 Acknowledgments

Jagdish Kumar acknowledges CSIR for providing financial assistance in the form of SRF. JK also acknowledges Dr. Jiji Pulikkotil for many fruitful discussions and consistent encouragement. SA would like to thank NPL for the J.C. Bose Fellowship. We thank Prof. H. Ebert for allowing us to use SPRKKR code.

**Figure Captions**

**Figure 1:** c-axis view of different antiferromagnetic states discussed for Fe based superconductors; (a) Checker board (CH), (b) Single stripes (SS) and (c) Double stripes (SS). Red and blue spheres represent Fe atoms with up and down spins respectively and yellow spheres correspond to chalcogen atoms.

**Figure 2:** Magnetic energy and moment on Fe as function of $h_{Se}$, studied for FeSe using (a) LDA and (b) GGA. Dashed vertical lines as indicated represent optimized ($h_{opt}$) and experimental ($h_{expt}$) value of $h_{Se}$. The background colors correspond to the magnetic ground state. Pink, yellow, grey and pale yellow correspond respectively to NM, SS, DS and FM.

**Figure 3:** Magnetic energy and magnetic moment on Fe as function of $h_{Te}$ for FeTe, (a) LDA and (b) GGA. Dashed vertical lines represent optimized and experimental value of $h_{Te}$. The background colors correspond to respective magnetic ground state as mentioned in figure 2.

**Figure 4:** (a) Electronic density of states at Fermi level and (b) Value of Stoner condition $I \times N(E_F)$ as function of chalcogen height above Fe layers for FeSe and FeTe. The value of *I* is taken from Janak [26].

**Figure 5:** Electronic band structure and corresponding FS of FeSe calculated for different values of $h_{Se}$. (a) 2.2939a.u. (b) 2.51a.u. (Optimized value), (c) 2.7861a.u. (Experimental value) and (d) 3.2854a.u. (*h* for FeTe)

**Figure 6:** Band structure (Bloch spectral function) calculated along different directions in Brillouin zone for FeTe$_x$Se$_{1-x}$ (a)-(e) and (f) the colour map used for all plots (see text).

**Figure 7:** Fermi surface plots along different planes in Brillouin zone for lattice parameters of (a) FeSe, (b) FeSe$_{0.5}$Te$_{0.5}$ and (c) FeTe. Figure (d) represents colour map for numerical values of BSF (see text).

**Table 1:** Structure parameters and results of force optimization done in different magnetic states for FeSe, FeTe$_{0.5}$Se$_{0.5}$ and FeTe. All the distances are in atomic units (Bohr).

|  | **Lattice Parameters in atomic units (Bohr)** |  | **Chalcogen height ($h_{Se}/h_{Te}$)** | | | | |
|---|---|---|---|---|---|---|---|
|  | a | c |  | NM | SS | DS | Expt. |
| FeSe | 7.1145 | 10.4271 | $h_{Se}$ | 2.5159 | 2.6741 | 2.7652 | 2.7861[14] |
| FeTe$_{0.5}$Se$_{0.5}$ | 7.1670 | 11.3667 | $h_{Se}$ | 2.4912 | 2.6575 | 2.7846 | 2.7932[43] |
|  |  |  | $h_{Te}$ | 3.0117 | 3.2147 | 3.3082 | 3.2460[43] |
| FeTe | 7.2203 | 11.8764 | $h_{Te}$ | 2.9774 | 3.2128 | 3.2960 | 3.3515[35] |

**Table 2:** Structural and electronic properties for FeTe$_x$Se$_{1-x}$ system. Lattice parameters are taken from experimental data. Values in brackets are experimental values. The average value of chalcogen height or bond distance is calculated by taking concentration average of their individual values. Fe-Se and Fe-Te indicates bond distances of Fe with Se and Te respectively. All the distances are in atomic units (Bohr). $N(E_F)$ is in states/eV/fu for both spins.

| | FeSe | FeTe$_{0.25}$Se$_{0.75}$ | FeTe$_{0.5}$Se$_{0.5}$ | FeTe$_{0.75}$Se$_{0.25}$ | FeTe |
|---|---|---|---|---|---|
| $a$ | 7.1145 | 7.1786 | 7.1700 | 7.2105 | 7.2203 |
| $c$ | 10.4271 | 11.2283 | 11.3667 | 11.7725 | 11.8764 |
| $z_{Se}$ | 0.2413 (0.2672)[14] | 0.2211 | 0.2192 (0.2468)[43] | 0.2065 | --- |
| $z_{Te}$ | --- | 0.2690 | 0.2649 (0.2868)[43] | 0.2550 | 0.2775 (0.2822)[35] |
| $z_{av}$ | 0.2413 | 0.2331 | 0.2420 (0.2778)[34] | 0.2429 | 0.2775 |
| $h_{Se}$ | 2.5159 (2.7861)[14] | 2.4826 | 2.4920 (2.7932)[43] | 2.4310 | --- |
| $h_{Te}$ | --- | 3.0204 | 3.0117 (3.2460)[43] | 3.0020 | 3.2959 (3.3515)[35] |
| $h_{av}$ | 2.5159 (2.7861)[14] | 2.6202 | 2.7513 (3.1577)[34] | 2.8572 | 3.2959 (3.3515)[35] |
| Fe-Se | 4.3571 (4.5185)[14] | 4.3662 | 4.3648 (4.5489)[43] | 4.3574 | --- |
| Fe-Te | --- | 4.6911 | 4.6810 (4.8393)[43] | 4.6869 | 4.8884 (4.9261)[35] |
| $N(E_F)$ | 1.41 | 1.48 | 1.56 | 1.61 | 2.48 {1.78[†]} |
| $\gamma_{calc.}$ (mJmol$^{-1}$K$^{-2}$) | 3.32 | 3.49 | 3.67 | 3.79 | 5.84 {4.05[†]} |
| $\gamma_{expt.}$ (mJmol$^{-1}$K$^{-2}$) | 9.17[1] | -- | ~23[48] | ~55* | 34[41] |
| $\lambda$ | 1.76 | -- | 5.26 | 13.5 | 4.82 |

\* For FeTe$_{0.70}$Se$_{0.30}$

[†] Obtained using NM optimized $h_{Te}$

**Figure 1(a)-(c)**

(a)

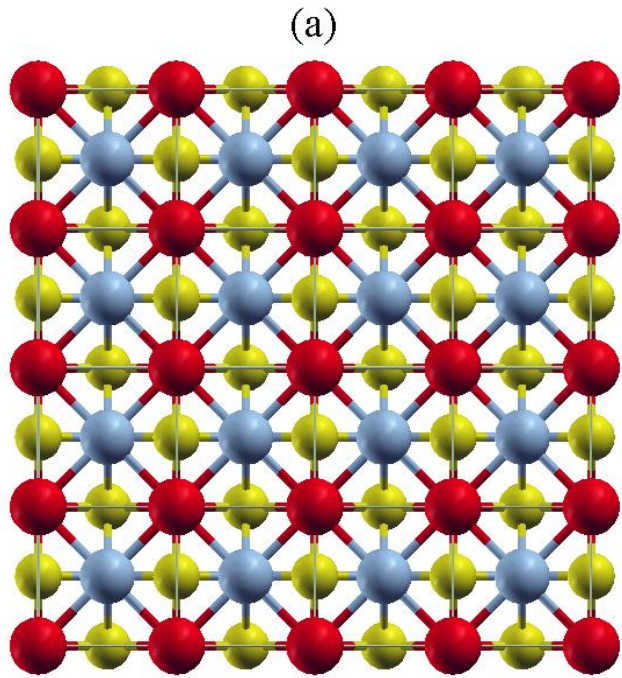

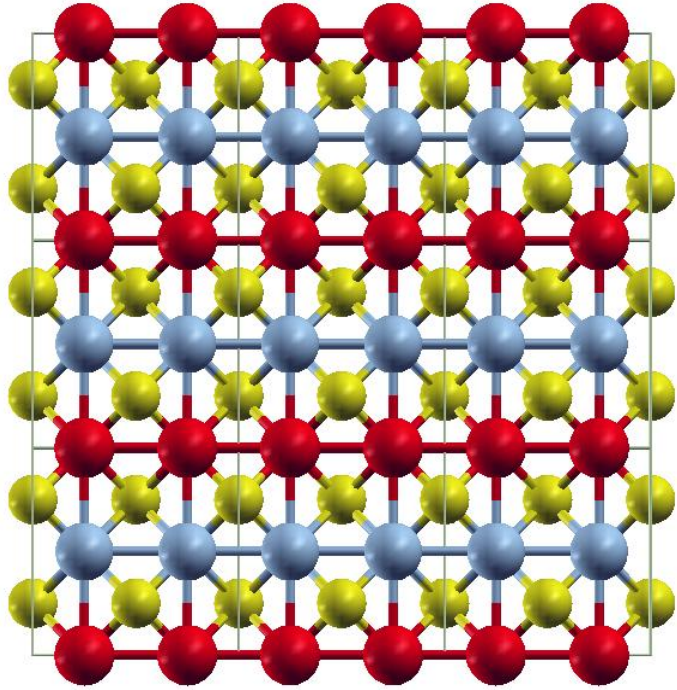

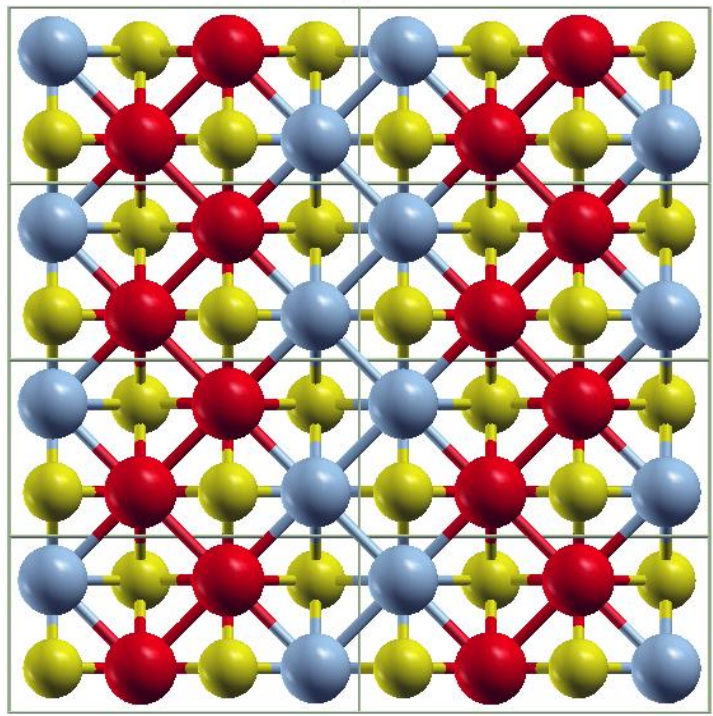

Figure 2(a)-(b)

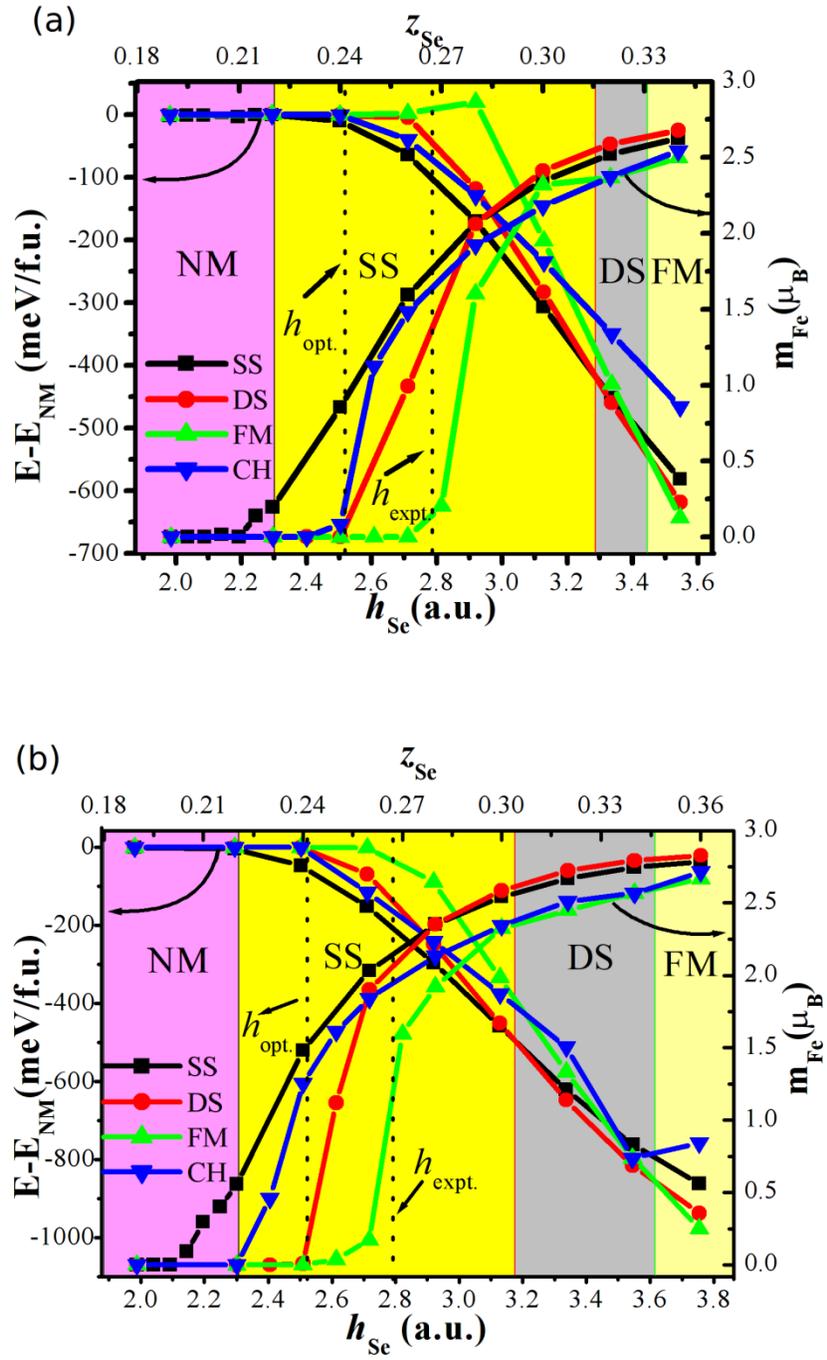

Figure 3(a)-(b)

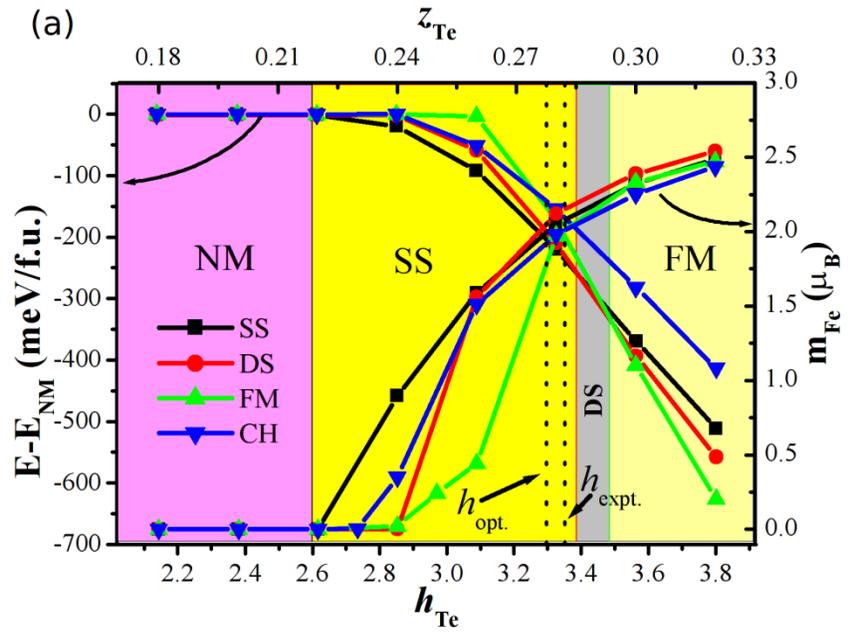

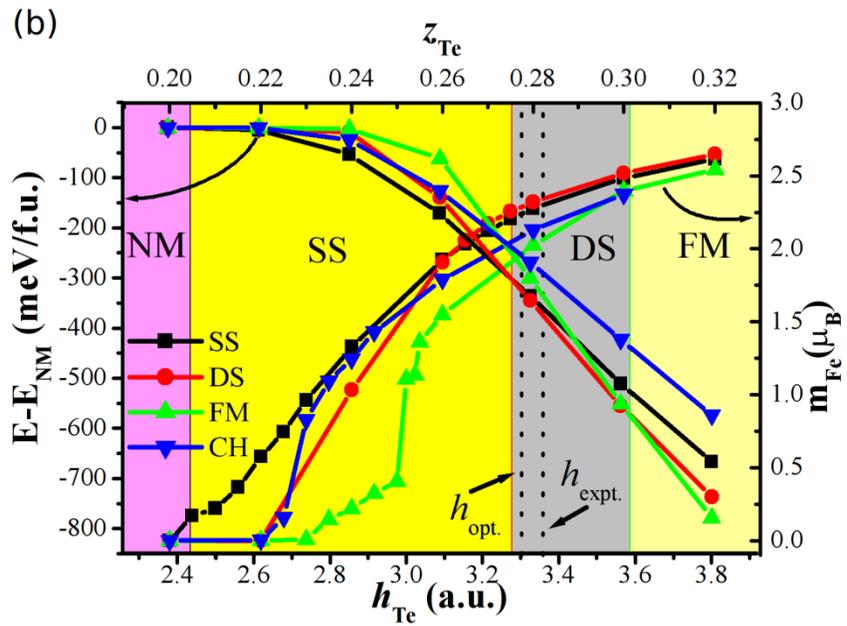

Figure 4(a)-(b)

(a)

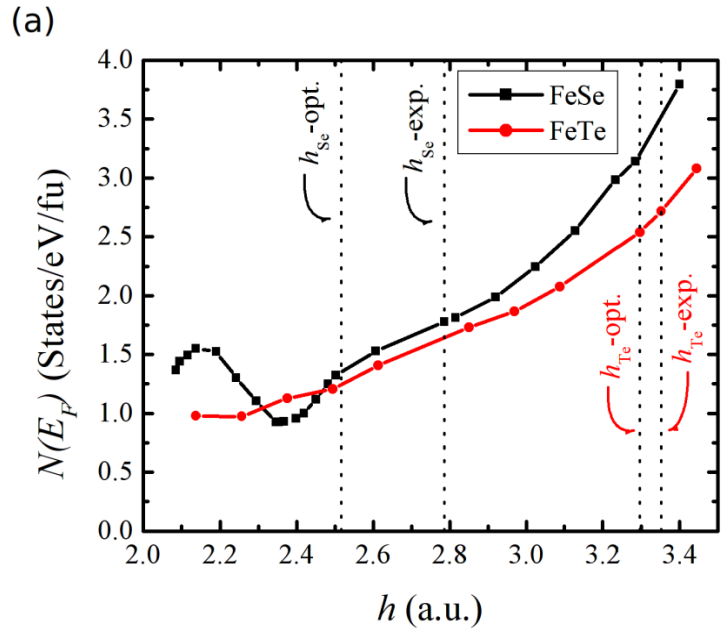

(b)

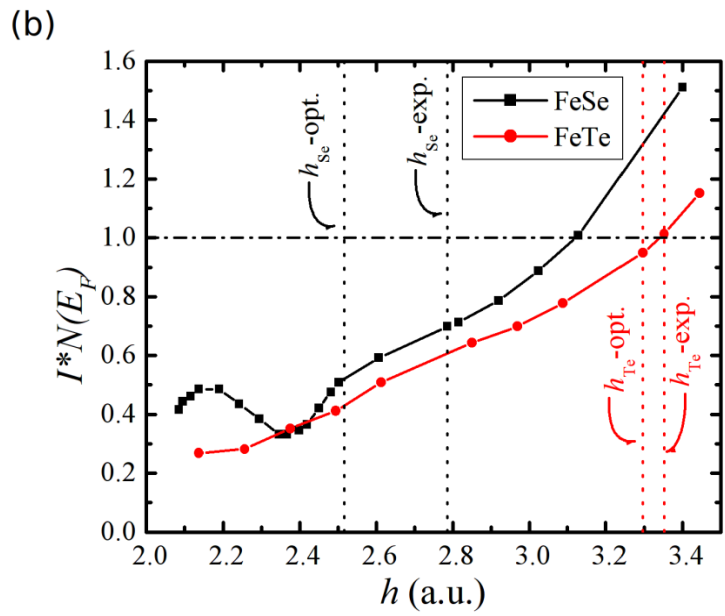

(a) z=0.2200, h=2.2939a.u.(1.2139Ang)

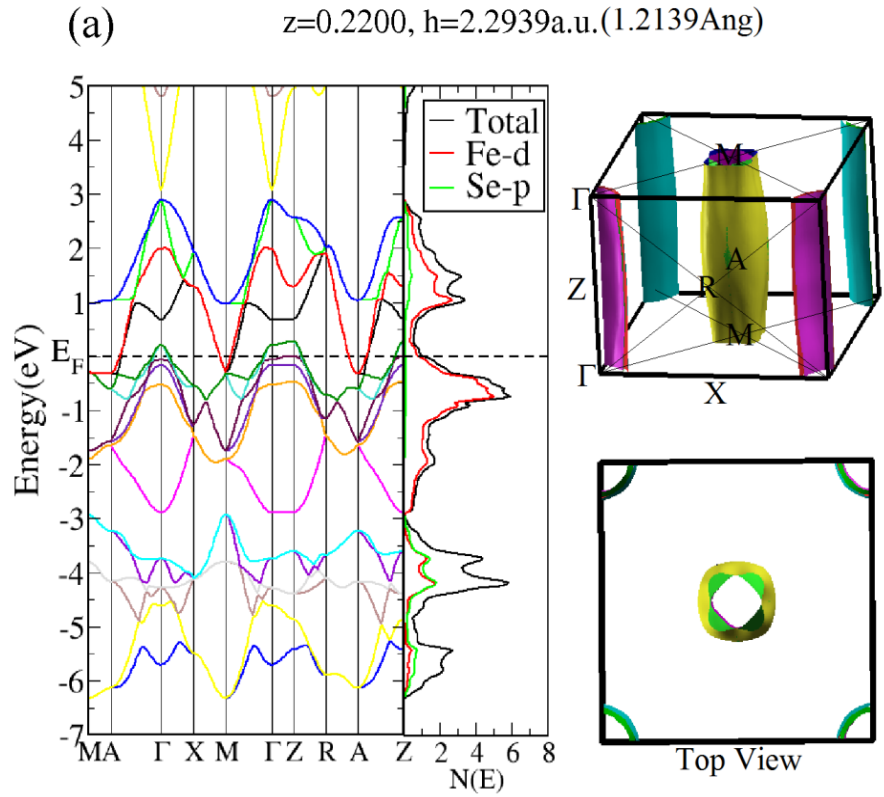

(b) FeSe at optimized z
z=0.2412, h=2.515a.u.(1.3309Ang)

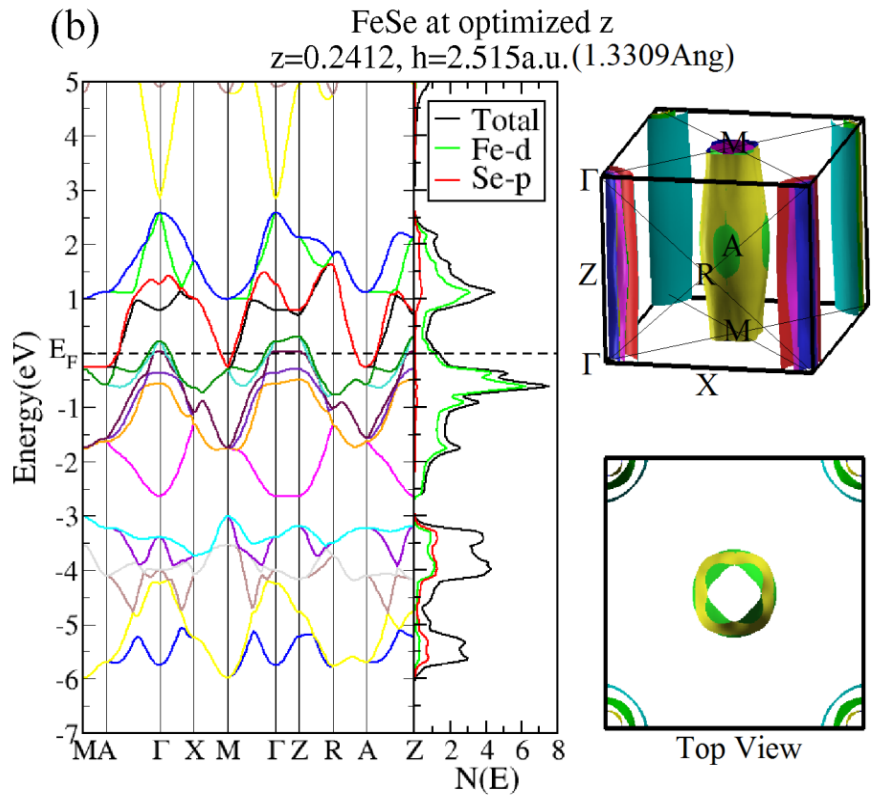

(c) FeSe for experimental z
z=0.2672, h=2.7861a.u.(1.4744Ang)

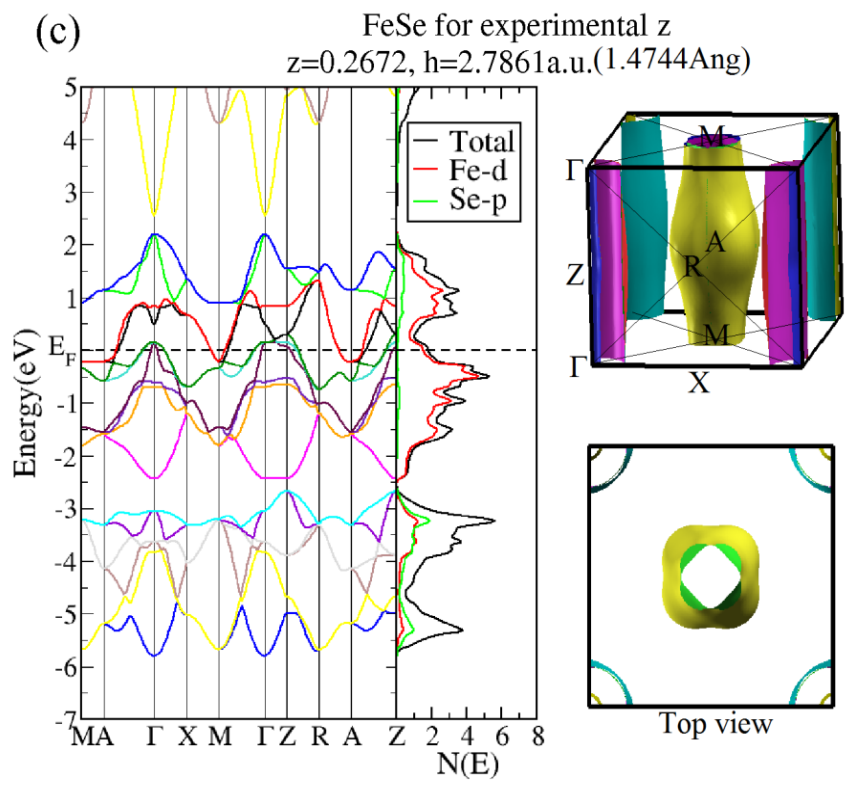

Top view

(d) z=0.315, h=3.2845a.u.(1.7382Ang)

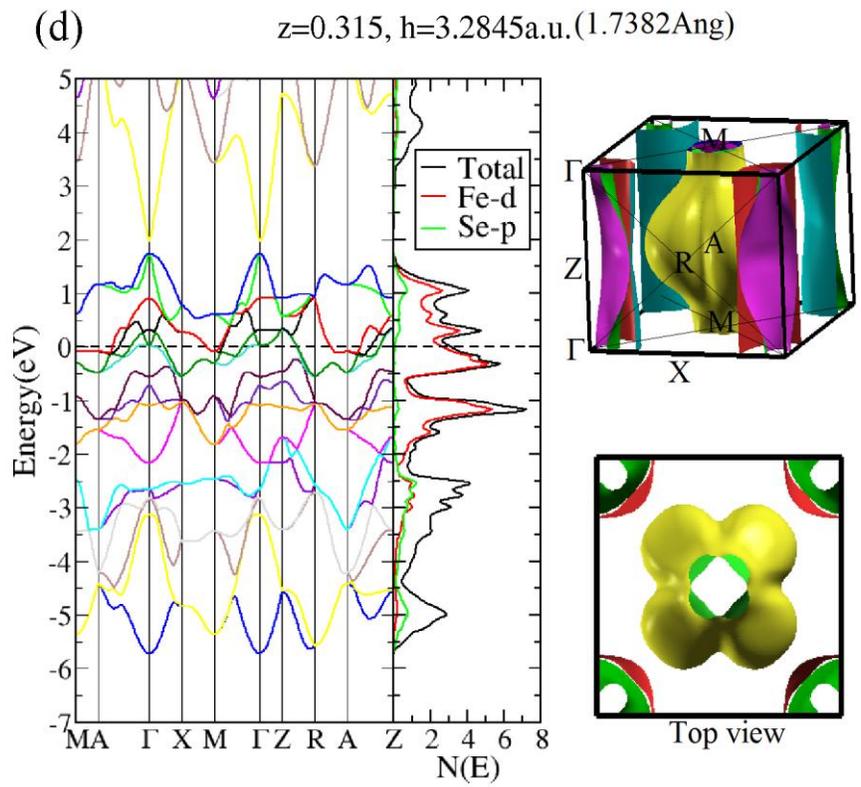

Top view

Figure 6(a)-(f)

(a)
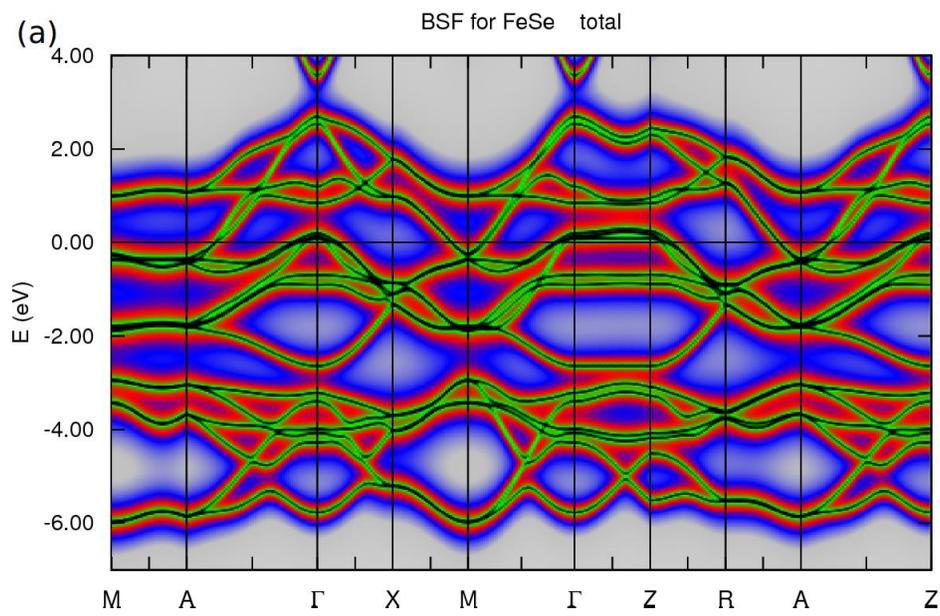

(b)
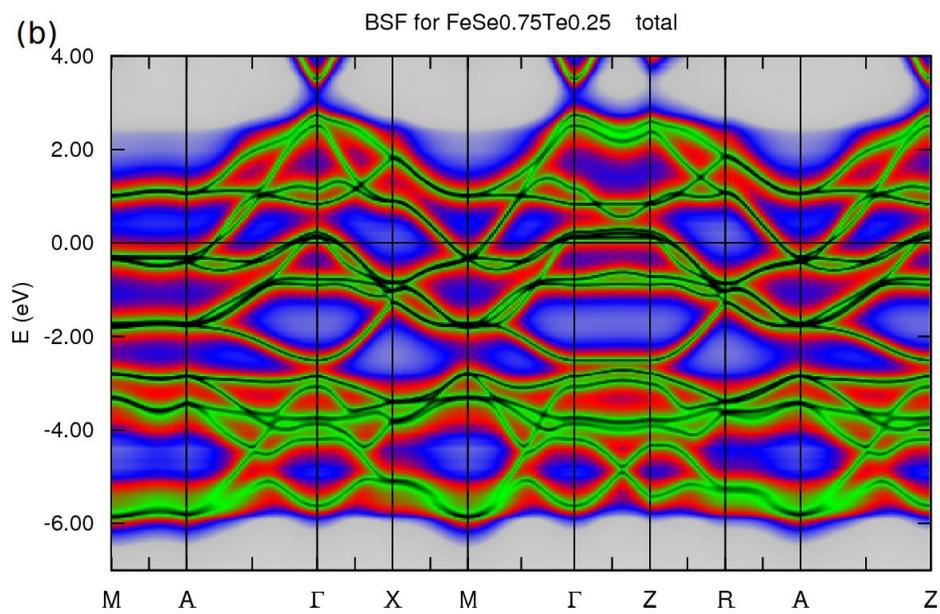

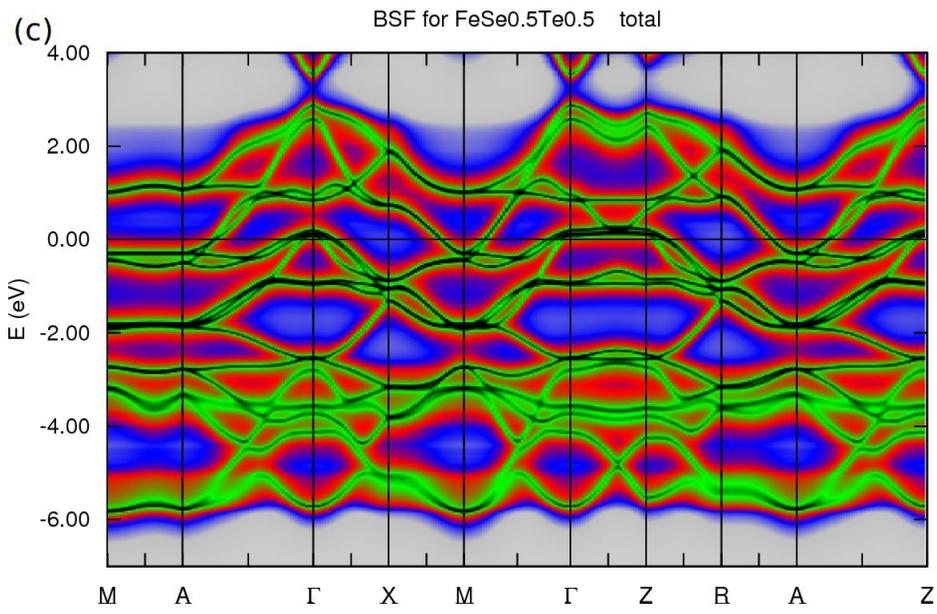

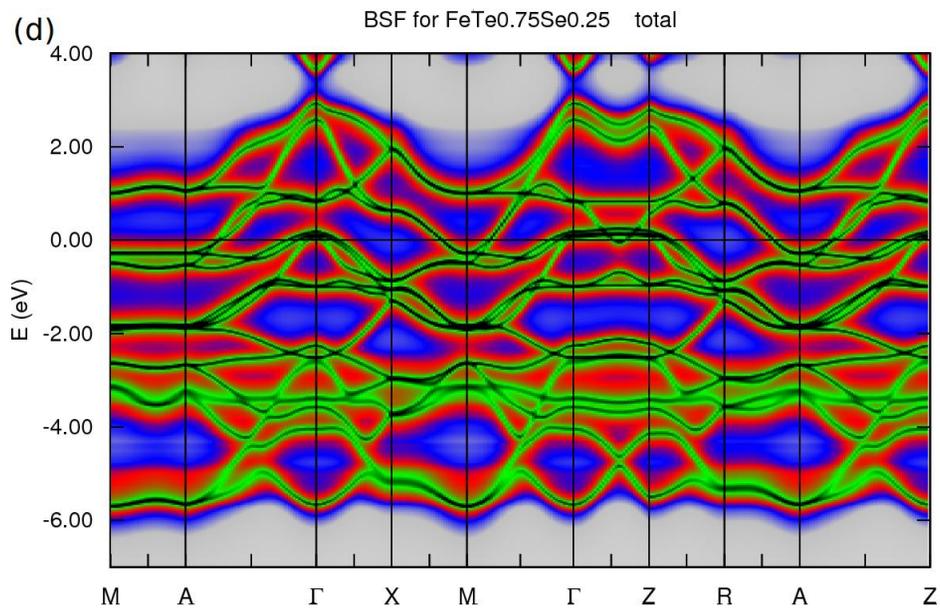

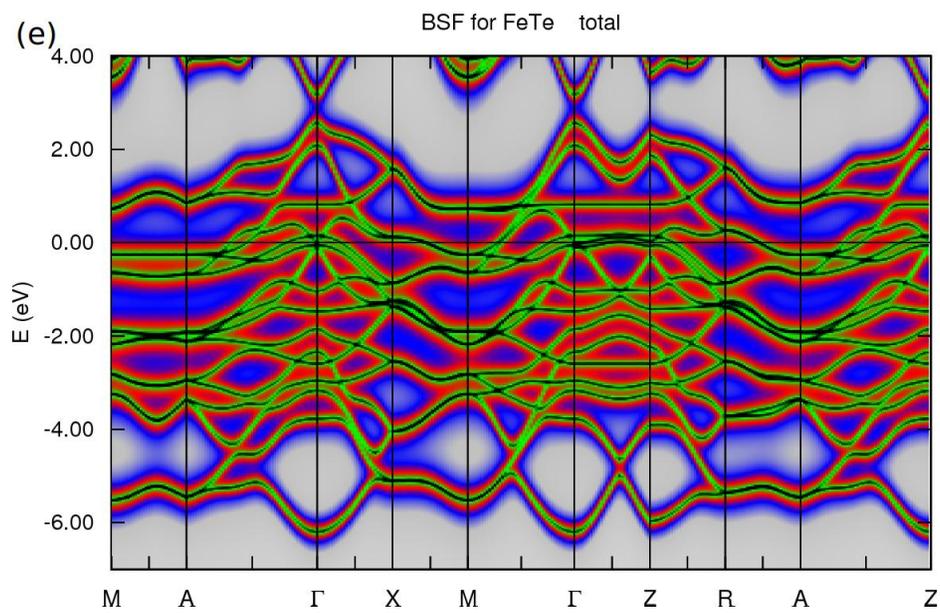

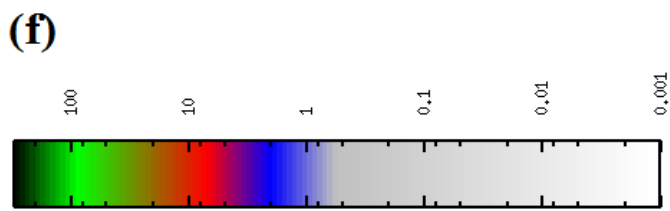

Figure 7(a)-(d)

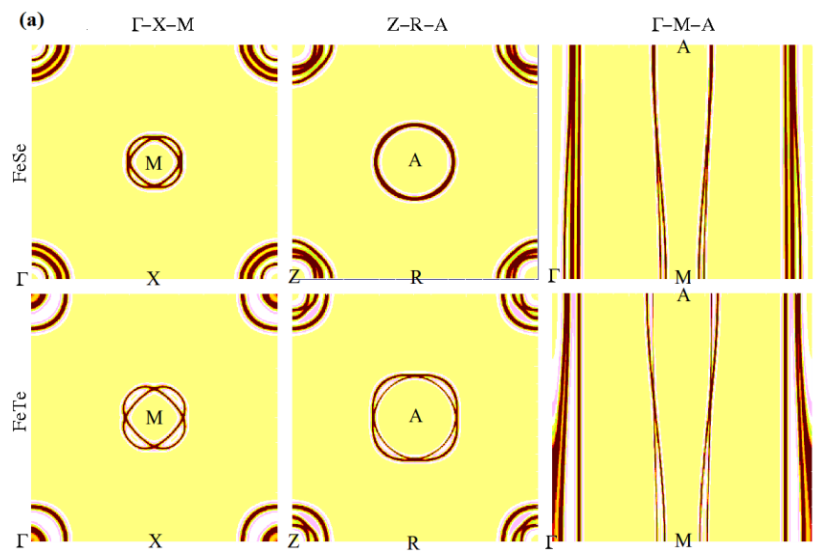

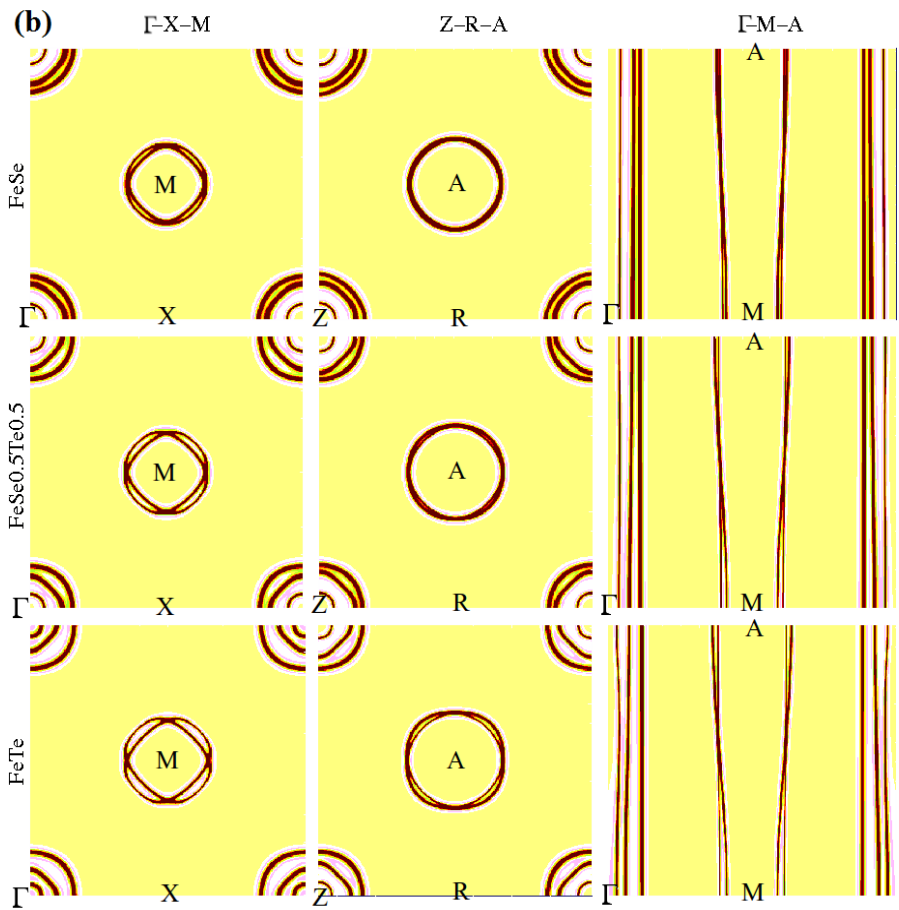

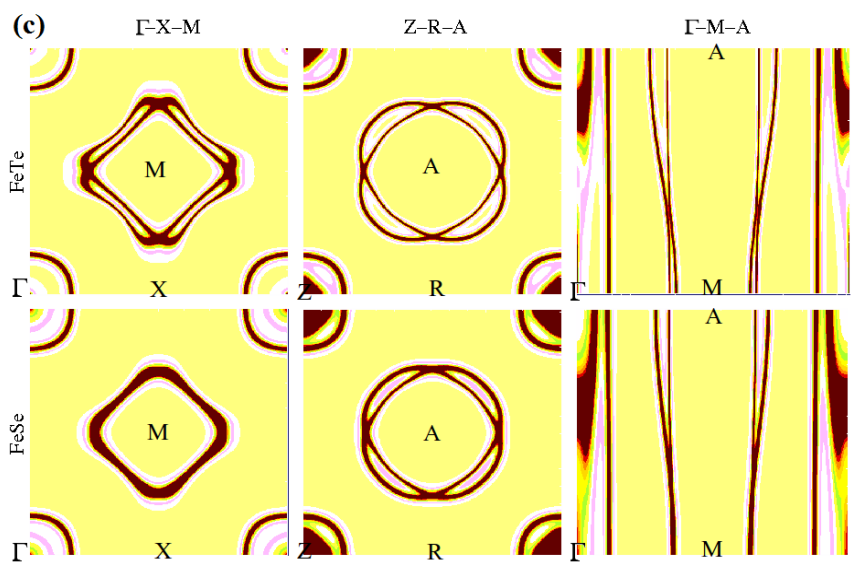

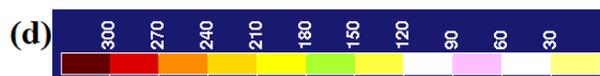